\documentclass{article}
\usepackage{spconf,amsmath,graphicx}
\usepackage{amssymb}
\usepackage{multirow}
\usepackage{booktabs}
\usepackage{xspace}
\ninept

\usepackage{xcolor}

\usepackage{makecell}

\def\x{{\mathbf x}}

\def\ours{{\it CHARM}\xspace}
\def\Ours{{\it CHARM}\xspace}
\def\CharmBase{{\it CHARM-base }}
\def\CharmCKV{{\it CHARM-CKV }}
\def\CharmCQ{{\it CHARM-CQ }}

\title{Learning from heterogeneous eeg signals\\ 
with differentiable channel reordering}
\name{Aaqib Saeed, David Grangier, Olivier Pietquin, Neil Zeghidour}
\address{Author Affiliation(s)}

\name{Aaqib Saeed$^1$\sthanks{This work was conducted while interning at Google.}, David Grangier$^2$, Olivier Pietquin$^2$, Neil Zeghidour$^2$}
\address{$^1$Eindhoven University of Technology, Eindhoven, The Netherlands \\ 
$^2$Google Research, Paris, France
}

\begin{document}
\maketitle

\begin{abstract}
We propose \Ours, a method for training a single neural network across inconsistent input channels. Our work is motivated by Electroencephalography (EEG), where data collection protocols from different headsets result in varying channel ordering and number, which limits the feasibility of transferring trained systems across datasets. Our approach builds upon attention mechanisms to estimate a latent reordering matrix from each input signal and map input channels to a canonical order. \Ours is differentiable and can be composed further with architectures expecting a consistent channel ordering to build end-to-end trainable classifiers. We perform experiments on four EEG classification datasets and demonstrate the efficacy of \Ours via simulated shuffling and masking of input channels. Moreover, our method improves the transfer of pre-trained representations between datasets collected with different protocols.
\end{abstract}
\section{Introduction}
\label{sec:intro}
Electroencephalography (EEG) is the measurement of the brain's electrical activity, which informs about neural functions and related physiological manifestations~\cite{kennett2012modern}. It is generally collected along the scalp in a non-invasive way for a wide array of tasks, including for clinical purposes and Brain-Computer Interface (BCI) systems. Automatic classification of EEG signals with machine learning has been widely adopted to study, diagnose, and treat neurological disorders such as seizures, epilepsy, Alzheimer's, and sleep-related problems~\cite{svanborg1996eeg, korkalainen2019accurate, supratak2017deepsleepnet}. In BCI tasks, EEG is used to capture motor imagery signals and recognize user's intents~\cite{pfurtscheller2001motor} and event-related potential~\cite{galloway1990human}. Likewise, it is also used to estimate mental workload or task complexity for monitoring cognitive stress and performance~\cite{li2016deep}.  

Over the last years, automatic EEG classification has moved from using hand-crafted features~\cite{shoeb2010application, craik2019deep} towards learning high-level representations from raw EEG signals with deep neural networks~\cite{wulsin2011modeling,ren2014convolutional}. In particular, Convolutional Neural Networks (CNNs) have become the standard architecture to process EEG signals and have been used for many tasks including motor imagery \cite{tabar2016novel, schirrmeister2017deep, lawhern2018eegnet}, seizure prediction \cite{thodoroff2016learning, zhou2018epileptic}, Parkinson diagnosis \cite{oh2018deep} and sleep stage scoring~\cite{supratak2017deepsleepnet}. Nevertheless, EEG measurements remain notoriously subject to intra- and inter-subject variability, which makes generalization particularly challenging %
and led to numerous works focusing on the reduction of this generalization gap \cite{fahimi2019inter,raghu2020eeg,wu2014transfer}. A less explored problem is the variability due to measuring devices: different EEG headsets have a varying number of electrodes (from a few to dozens) and different electrical specifications \cite{wu2016switching}. Moreover, it is not rare that headset malfunctions lead to noisy or even missing channels. Consequently, available EEG datasets are heterogeneous, and the majority of them are very small. 

Scaling EEG training data seems, therefore, only feasible by aggregating heterogeneous datasets. This requires devising novel classification methods that are robust to permuted and missing channels since classical CNNs assume a fixed number of input channels, ordered consistently across data examples. With this objective, we introduce a new framework for training a single CNN across varying EEG collections that can differ both in number and location of electrodes. Our CHAnnel Reordering Module (CHARM) ingests multichannel EEG signals, identifies the location of each channel from their content, and remaps them to a canonical ordered set. After this remapping, the channels are ordered consistently and can be further processed by a standard neural network, regardless of the actual variations in the input data. We evaluate CHARM on three tasks: seizure classification, detection of abnormal EEGs and detection of artifacts (e.g. eye movement). We show that CHARM is significantly more robust to missing and permuted channels than a standard CNN. We also introduce a data augmentation technique that further improves the robustness of the model. Moreover, we show for the first time that pre-training representations of EEG on a large dataset transfers to another, smaller dataset collected with a different headset.

\begin{figure*}[!htbp]
\centering
\includegraphics[width=\textwidth]{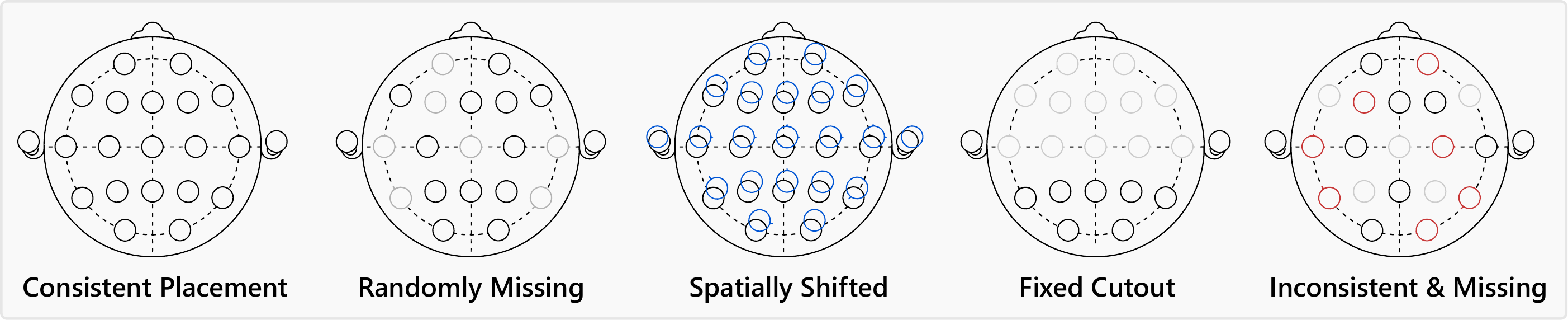}
\caption{Illustration of various forms of inconsistencies that can arise in EEG recordings.}
\label{fig:overview_placement}
\end{figure*}

\section{Methodology}
\label{sec:method}
Our proposal is a differentiable reordering module that maps inputs with inconsistent channels to a fixed canonical order. It can be composed with further modules expecting consistent channel placement to be trained end-to-end on data without channel ordering information. As inputs, we consider EEG signals with an unknown channel ordering and potentially missing channels, (Figure~\ref{fig:overview_placement}). Our module takes these channels and reorders them to a canonical, consistent order prior to further processing. Precisely, our reordering module outputs a soft reordering matrix $p$.

\begin{figure*}[!htbp]
\centering
\includegraphics[width=6.2in]{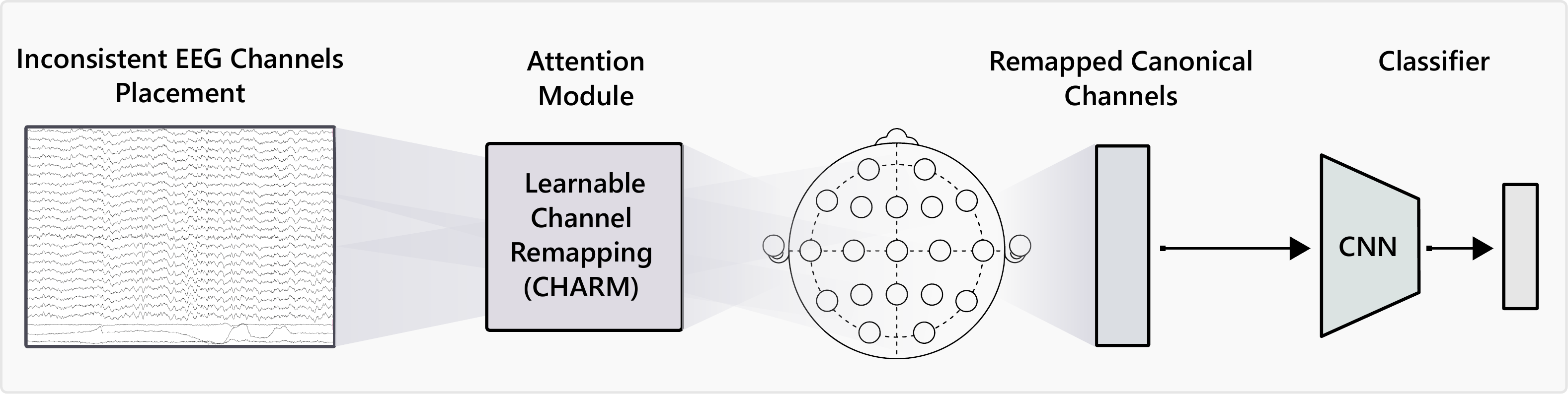}
\caption{Overview of CHARM with a $1$D convolutional classifier for EEG classification tasks.}
\label{fig:overview_approach}
\end{figure*}

The input signal $\x \in \mathbb{R}^{N \times T}$ is a recording over $N$ channels for a duration $T$. Considering $M$ canonical channels, the reordering matrix $p(\x)$ is an $N \times M$ matrix. Precisely, each canonical output is estimated as a weighted sum of the input channels, i.e.,
\begin{align}
     \hat{\mathbf{x}}_{i, t} = \sum_{j=1}^N p(\x)_{i, j} ~ \x_{j, t}\hspace{0.25em}, \quad i = 1, \ldots, M, \quad t = 1, \ldots, T.
\end{align}
$\hat{\x} \in \mathbb{R}^{M \times T}$ then serves as input to a standard neural network expecting a consistent input ordering across data samples. Since $p$ is differentiable, the reordering module parameters can be learned jointly with the rest of the architecture (Figure~\ref{fig:overview_approach}). Training optimizes the cross-entropy classification loss with no extra supervision on the channel order. 

\subsection{Learnable Channel Remapping}

Using \Ours as the first layers of the model allows us to train a single deep architecture over different EEG recording headsets. 
We consider three variants of our reordering method.

\subsubsection{Convolutional Reordering}
\label{sec:base_model}

\CharmBase{} represents the signal of each channel as a vector,
\begin{equation}
h_{i} = m^{\rm conv}(\x_{i, :})
\end{equation}
where $m^{\rm conv}$ composes a 1-D convolution layer with $d$ filters and an aggregation operation (global max-pooling) to map a single-channel temporal signal into a fixed dimensional vector of dimension $d$. Since this step convolves channels independently, its predictions are invariant to a reordering of the input channels.

Each vector is then compared to learned embeddings of dimension $d$ that represent each of the $M$ canonical channels, $c \in \mathbb{R}^{M \times d}$, yielding the matrix $p$ for channel remapping,
\begin{equation}
\label{eq:base_perm}
p_{i, j} = {\rm softreorder}(c, h)_{i,j} = \frac{\exp(c_{i} \cdot h_{j})}{\sum_{j'} \exp(c_{i} \cdot h_{j'})}.
\end{equation}
\subsubsection{Attentive Reordering with Canonical Keys and Values}

\CharmCKV{} builds upon residual attention~\cite{vaswani2017attention}. We build a {\it query vector} representing each input channel as
\mbox{$
q_{i} = m^{\rm conv}(\x_{i, :}).
\nonumber
$}
Each query vector {\it attends} over canonical channels, i.e. each input channel query is mapped to a weighted sum of canonical channel {\it value vectors} according to their similarity to canonical {\it key vectors},
\begin{align}
h_{i} = \sum_j a_{i,j} v_j
\quad
\textrm{where}
\quad
a_{i,j} = \frac{\exp(q_{i} \cdot k_{j})}{\sum_{j'} \exp(q_{i} \cdot k_{j'})},
\end{align}
and $k$, $v$ are key and value embeddings representing the canonical channels. These layers can be stacked after a residual connection, 
\begin{align}
q^{l+1}_{i} = {\rm layernorm}(q^{l}_{i} + {\rm mlp}^l(h^l_{i}))
\end{align}
where $q^l, h^l$ are the query and attentive representations of layer $l$. 
${\rm layernorm}$ denotes a layer normalization module \cite{ba2016layer} and ${\rm mlp}$ denotes a multi-layer perceptron with a single hidden layer. We denote the residual 
attention compactly as 
\begin{align}
q^{l+1}_{i} = {\rm attn}(q^{l}_{i}, k, v)
\label{eq:attention}
\end{align}
At the last layer, we compute \mbox{$p = {\rm softreorder}(c, q^l).$}

Compared to our base convolutional model, this model can use many channel predictions to refine each individual prediction. For instance, it can be confident only for a few channels at the first layer and refine its decision for the other channels in the next layers.

\subsubsection{Attentive Reordering with Canonical Queries}

\CharmCQ{} reverses the role of canonical and input channels, using input keys and values and relying on canonical queries. Input keys and values result from an independent channel-wise convolution with $1$D filters,
\begin{align}
(k_{i}, v_{i}) = m^{\rm conv}(\x_{i, :})
\end{align}
and initial canonical queries $q$ are learned embeddings. Each query can be attended over to represent each {\it canonical channel} as a weighted sum of {\it input} values, as in Equation~\ref{eq:attention}. Several layers of attention can be stacked. Finally, 
\mbox{$p = {\rm softreorder}(q^l, k^l).$}

In our experiment, we found it beneficial to share keys and values across layers, i.e. $(k^l, v^l) = (k^0, v^0).$

\subsection{CMSAugment: shuffling and masking channels}
Data augmentation is another strategy to improve the generalization to inconsistent inputs. As an orthogonal contribution to our remapping module, we propose a \textsc{Channel Masking and Shuffling Augmentation} (CMSAugment) strategy. During training, CMSAugment first shuffles the channels and next samples a binary mask over channels to drop some channels entirely, with a uniform distribution from 0 masked channels to $N-1$. In Section \ref{sec:experiments}, we show how it significantly helps a simple CNN becoming robust to missing and inconsistent channel placements, with the best results being obtained by combining \ours and CMSAugment.

\subsection{Network Architecture Design and Implementation}

Our main architecture is a 1D-CNN that takes raw EEG signals as input and cascades four blocks of $1$D convolutions. Each block has a LayerNormalization~\cite{ba2016layer} and PReLU \cite{he2015delving} as activation along with a max-pooling layer for downsampling. We use a kernel size of $8$ with stride $1$ and pooling size of $2$ and stride $2$ with $256$ feature maps for the first three blocks, and $512$ in the last one. To aggregate the features, we use global max-pooling , which then feeds into a single linear classification layer. We apply L$2$ regularization with a rate of $10^{-4}$ on all the layers but the last.
Given an input sequence, we first apply an InstanceNormalization~\cite{ulyanov2016instance}, which normalizes each channel independently. Then our baseline model (Baseline) passes the output through the 1D-CNN directly. On the other hand, \ours first passes the waveforms through the remapping module, which then feeds into the 1D-CNN.

\begin{table*}[!t]
\centering
\caption{Test accuracy ($\pm$ std) averaged over $10$-folds for model generalization to shuffled and masked input channels. The entries with Noisy-$n$\% represent the results when a fixed ratio of $n\%$ of the channels are masked after shuffling.}
\vspace{0.09cm}
\label{tab:main}
\footnotesize
\begin{tabular}{@{}lcclccccc@{}}
\toprule
\textbf{Dataset} & \textbf{Channels} & \textbf{Classes} & \textbf{Method} & \textbf{Clean}  & \textbf{Noisy}  & \textbf{Noisy-25\%} & \textbf{Noisy-50\%} & \textbf{Noisy-75\%} \\ \midrule
\multirow{4}{*}{TUH Abnormal}  & \multirow{4}{*}{22} & \multirow{4}{*}{2} & Baseline        & \textbf{0.830   $\pm$ 0.030} & 0.566   $\pm$ 0.025 & 0.581   $\pm$ 0.025     & 0.589   $\pm$ 0.027     & 0.548   $\pm$ 0.028     \\
 & & & \CharmBase{} & 0.766   $\pm$ 0.016  & 0.742   $\pm$ 0.014 & 0.760   $\pm$ 0.015 & 0.743   $\pm$ 0.016 & 0.731   $\pm$ 0.014  \\
 & & & \CharmCKV{}       & 0.772   $\pm$ 0.019  & \textbf{0.751   $\pm$ 0.011} & \textbf{0.767   $\pm$ 0.013} & \textbf{0.756   $\pm$ 0.013} & \textbf{0.747   $\pm$ 0.015}  \\
 & & & \CharmCQ{}   & 0.751   $\pm$ 0.014  & 0.743   $\pm$ 0.022 & 0.744   $\pm$ 0.026 & 0.744   $\pm$ 0.024 & 0.741   $\pm$ 0.028  \\ \midrule
\multirow{4}{*}{TUH Artifact} & \multirow{4}{*}{19} & \multirow{4}{*}{6} & Baseline        & \textbf{0.711   $\pm$ 0.009} & 0.243   $\pm$ 0.016 & 0.245   $\pm$ 0.026     & 0.176   $\pm$ 0.018     & 0.263   $\pm$ 0.022     \\
 & & & \CharmBase{} & 0.618   $\pm$ 0.011  & 0.514   $\pm$ 0.028 & \textbf{0.566   $\pm$ 0.029} & 0.517   $\pm$ 0.028 & 0.466   $\pm$ 0.028  \\
 & & & \CharmCKV{}       & 0.628   $\pm$ 0.016  & 0.481   $\pm$ 0.028 & 0.521   $\pm$ 0.034 & 0.491   $\pm$ 0.026 & 0.452   $\pm$ 0.019 \\
 & & & \CharmCQ{}   & 0.607   $\pm$ 0.009  & \textbf{0.524   $\pm$ 0.044} & 0.538   $\pm$ 0.043 & \textbf{0.531   $\pm$ 0.046} & \textbf{0.505   $\pm$ 0.043}  \\ \midrule
\multirow{4}{*}{TUH Seizure}  & \multirow{4}{*}{21} & \multirow{4}{*}{5} & Baseline        & \textbf{0.950   $\pm$ 0.010} & 0.289   $\pm$ 0.040 & 0.368   $\pm$ 0.037     & 0.297   $\pm$ 0.052     & 0.171   $\pm$ 0.046     \\
 & & & \CharmBase{} & 0.906   $\pm$ 0.021 & 0.663   $\pm$ 0.015 & 0.818   $\pm$ 0.026 & 0.693   $\pm$ 0.034 & 0.502   $\pm$ 0.027  \\
 & & & \CharmCKV{}       & 0.912   $\pm$ 0.027  & 0.713   $\pm$ 0.030 & 0.842   $\pm$ 0.041 & 0.756   $\pm$ 0.043 & 0.591   $\pm$ 0.046  \\
 & & & \CharmCQ{}   & 0.890   $\pm$ 0.021  & \textbf{0.770 $\pm$ 0.041} & \textbf{0.857 $\pm$ 0.028} & \textbf{0.808 $\pm$ 0.045} & \textbf{0.704 $\pm$ 0.058}  \\ \midrule
\multirow{4}{*}{CHB-MIT}   & \multirow{4}{*}{17}  & \multirow{4}{*}{2}  & Baseline        & \textbf{0.658 $\pm$ 0.009} & 0.371   $\pm$ 0.014 & 0.296   $\pm$ 0.014     & 0.363   $\pm$ 0.012     & 0.439   $\pm$ 0.029     \\
 & & & \CharmBase{} & 0.554   $\pm$ 0.006  & 0.504   $\pm$ 0.010 & 0.523   $\pm$ 0.012 & 0.503   $\pm$ 0.013 & 0.487   $\pm$ 0.016  \\
 & & & \CharmCKV{}    & 0.562   $\pm$ 0.009  & 0.518   $\pm$ 0.011 & 0.543   $\pm$ 0.009 & 0.529   $\pm$ 0.007 & 0.504   $\pm$ 0.009  \\
 & & & \CharmCQ{}  & 0.576   $\pm$ 0.009  & \textbf{0.541  $\pm$ 0.009} & \textbf{0.560 $\pm$ 0.010} & \textbf{0.550 $\pm$ 0.010} & \textbf{0.530 $\pm$ 0.014}  \\ \bottomrule
\end{tabular}
\end{table*}

\subsubsection{Channel Remapping Networks}

\Ours uses $24$ canonical channels, regardless of the actual number of channels in an input sequence (from 17 to 22 in our experiments). We represent each canonical channel with an embedding of dimension $d = 32$. \CharmBase{} contains three convolutional layers with $32$ feature maps and a filter size of $8$ with a stride of $1$. The residual attentive modules are inspired by Transformers~\cite{vaswani2017attention}. Queries, keys and values have a dimension of $32$, and the ${\rm mlp}$ has a single hidden layer of dimension $64$. For all models, we apply $L1$ regularization onto $p$ with a weight of $10^{-4}$ to promote sparse reordering matrices.

\subsubsection{Training Details}

We train on $500$-sample windows dynamically sampled from an entire EEG sequence. \Ours is trained jointly with the 1D-CNN to minimize a categorical cross-entropy loss, using ADAM \cite{kingma2014adam} with a learning rate of $10^{-4}$ and a batch size of $64$ for $100$ epochs. For imbalanced datasets (Section~\ref{sec:data}), we use a weighted cross-entropy loss to minimize error across rare and frequent classes equally.

\section{Experiments}
\label{sec:experiments}
\label{sec:data}
Our evaluation focuses on the Temple University Hospital EEG Corpus (TUH)~\cite{obeid2016temple} and CHB-MIT dataset~\cite{shoeb2009application}. The TUH corpus is the most extensive publicly available corpus with over $15000$ subjects; it comprises several datasets analyzed and annotated by expert clinicians where the majority of EEG is sampled at $250$Hz~\cite{obeid2016temple}. The CHB-MIT contains intractable seizures collected from $23$ pediatric subjects at a sampling rate of $256$Hz. Here, we focus on the tasks of recognizing abnormal EEG (TUH Abnormal), detection of artifacts such as eye movement or chewing (TUH artifacts) as well as determining the presence and type of seizures (TUH Seizure, CHB-MIT). Each dataset has a different number of EEG channels, as shown in Table \ref{tab:main}. We employ a $10$-folds stratified cross-validation technique for assessing the model performance. Our evaluation metric is the accuracy averaged over ten folds, weighted for imbalanced datasets (TUH Artifacts, TUH Seizure) to account for minority classes.  

\begin{table*}[!htbp]
\centering
\caption{Performance when masking half of the brain, along a vertical or horizontal axis.}
\vspace{0.09cm}
\label{tab:structure_mask}
\footnotesize
\begin{tabular}{@{}llccccc@{}}
\toprule
\multirow{2}{*}{\textbf{Augmentation}} &
  \multirow{2}{*}{\textbf{Method}} &
  \multirow{2}{*}{\textbf{Clean}} &
  \multicolumn{2}{c}{\textbf{Horizontal}} &
  \multicolumn{2}{c}{\textbf{Vertical}} \\
&                 &               & \textbf{$\text{Group}_A$} & \textbf{$\text{Group}_{B}$} & \textbf{$\text{Group}_A$} & \textbf{$\text{Group}_B$} \\ \midrule
\multirow{3}{*}{None} & Baseline        & \textbf{0.951 $\pm$ 0.007}   & 0.631 $\pm$ 0.050       & 0.430 $\pm$ 0.043   & 0.486 $\pm$ 0.028     & 0.586 $\pm$ 0.052     \\
& \CharmCKV{}       & 0.900 $\pm$ 0.034   & 0.790 $\pm$ 0.032     & 0.600 $\pm$ 0.023     & 0.683 $\pm$ 0.033     & 0.762 $\pm$ 0.041     \\
& \CharmCQ{}   & 0.899 $\pm$ 0.018   & \textbf{0.839 $\pm$ 0.030}     & \textbf{0.707 $\pm$ 0.056}  & \textbf{0.751 $\pm$ 0.040}   & \textbf{0.824 $\pm$ 0.028}     \\ \midrule
\multirow{3}{*}{CMSAugment} & Baseline & \textbf{0.873 $\pm$ 0.024} & 0.823 $\pm$ 0.037     & 0.762 $\pm$ 0.02     & 0.779 $\pm$ 0.036     & 0.783  $\pm$ 0.03 \\
& \CharmCKV{} & 0.829 $\pm$ 0.038   & \textbf{0.850 $\pm$ 0.029} & \textbf{0.778 $\pm$ 0.034} & \textbf{0.794 $\pm$ 0.037} & \textbf{0.853 $\pm$ 0.022}     \\
& \CharmCQ{}   & 0.734 $\pm$ 0.224   & 0.750 $\pm$ 0.217     & 0.702 $\pm$ 0.200     & 0.711 $\pm$ 0.203      & 0.739 $\pm$ 0.213     \\ \bottomrule
\end{tabular}
\end{table*}

\subsection{Generalizing to shuffled and masked channels}
\label{sec:shuffled_masked}
Table~\ref{tab:main} compares \ours to a baseline 1D-CNN when generalizing to noisy conditions. The models are trained on \textit{clean} (no masking, no shuffling) channels, but the evaluation is done under different forms of noise injection. To this end, the \textit{Noisy} entries in Table~\ref{tab:main} indicate performance when the test input channels are shuffled and $0$ to $N-1$ channels are uniformly masked. Similarly, Noisy-$n\%$ represents the results when a fixed ratio of $n\%$ channels are masked at random after shuffling. We first observe that when evaluating on clean inputs, the baseline model performs better. This can be explained by the fact that \Ours sees channels independently and cannot exploit ground-truth channel location, which is useful when training and evaluating on identical channels. On the other hand, \Ours-based remapping techniques perform significantly better in handling permuted and masked channels across all four datasets. Even when $50\%$ to $75\%$ channels are missing, our proposed approach maintains high accuracy. In particular, on TUH Seizure/Noisy-75\% \CharmCQ{} attains $0.704$ accuracy, against $0.171$ for the baseline.

\subsection{Performance in structured masking conditions} %
\label{sec:half_masked}

Table~\ref{tab:structure_mask} reports the performance on TUH Seizure when only a subset of the channels from a specific half of the brain is active, a more tangible setting than random masking. We split the electrodes along a vertical or horizontal axis and evaluate on each half separately ($\text{Group}_A$ and $\text{Group}_B$). We train the models either on clean inputs or with CMSAugment, and do not report results for \CharmBase{} since it performed worse than alternatives in previous experiments. %
When training without augmentation, \ours significantly outperforms the baseline in most cases, reaching its best results when combined with CMSAugment. Interestingly, the robustness of the baseline system significantly improves when trained with CMSAugment. This shows that, independently of \Ours, data augmentation is also a promising avenue for robust EEG classification.

\begin{table}[t]
\centering
\caption{Out of domain transfer results on CHB-MIT with a model pre-trained on TUH Seizure dataset. }
\footnotesize
\vspace{0.09cm}
\label{tab:transfer}
\begin{tabular}{@{}lcc@{}}
\toprule
\textbf{Method} & \textbf{Fixed}     & \textbf{Fine-tuned} \\ \midrule
Baseline (in-domain)   & \multicolumn{2}{c}{0.891 $\pm$ 0.038}      \\
Baseline (transfer)        & 0.757 $\pm$ 0.004    & \textbf{0.963 $\pm$ 0.015}     \\
\CharmCKV{}       & \textbf{0.805 $\pm$ 0.008}    & 0.942 $\pm$ 0.020     \\
\CharmCQ{}   & 0.795    $\pm$ 0.006 & 0.915 $\pm$ 0.026     \\ \bottomrule
\end{tabular}
\end{table}

\subsection{Transfer learning}
\label{sec:transfer}

So far, we assessed the performance individually for each task, simulating different headsets with random shuffling and masking. Now, we evaluate the proposed methods in handling inconsistent channels in a real cross-dataset transfer learning setting. In \cite{wu2016switching}, authors propose an algorithm to handle transfer between headsets, for a same subject, and using the common subset of channels shared between headsets. In contrast, \Ours{} does not require any knowledge about channel placement and exploits all channels of each headset. Moreover, we experiment in a more challenging setting: we transfer trained representations to a new headset, on new subjects.
We pre-train the models with clean inputs on the TUH Seizure dataset in a standard way and discard the classification head. We then reuse the other layers for learning either a linear classifier on-top of a fixed network or fine-tune it entirely on the CHB-MIT dataset. Importantly, as the number of channels in CHB-MIT is lower than TUH Seizure, we pad them with zero channels. In Table~\ref{tab:transfer}, we report the results of the baseline CNN along with channel remapping modules, where the in-domain baseline is the model directly trained on the CHB-MIT dataset, i.e., no transfer is performed. We observe that the representations learned with \ours transfer better than the baseline when freezing the transferred layers. This shows that our system allows for pre-training and transfer of embeddings across heterogeneous datasets. Interestingly, if we fine-tune the entire network, then the baseline (transfer) is able to relearn low-level representations and improves significantly over the in-domain baseline. To the best of our knowledge, this is the first time that a deep EEG classifier demonstrates out-of-domain transfer to a dataset recorded with a different headset over different subjects.

\section{Conclusion}
We introduce \Ours\!\!, a channel remapping model for training a single neural network across heterogeneous EEG data. Our model identifies the location of each channel from their content and remaps them to a canonical ordering by predicting a reordering matrix. This allows further processing by standard classifiers that expect a consistent channel ordering. Our differentiable reordering module leverages attention mechanisms and can be trained on data without information on the channel placement. We complement this model with a new data augmentation technique and demonstrate the efficiency of our approach over three EEG classification tasks, where
various types of headsets can result in inconsistent channel orderings and numbers. In particular, we successfully transfer parameters across datasets with different collection protocols. This is an important result since available EEG data are currently fragmented across a wide variety of heterogeneous datasets. We believe that the robustness of our method will pave the way to training a single model across large-scale collections of heterogeneous data. Moreover, our approach is general enough to benefit other domains with varying sensor placements and numbers, including weather modeling, seismic activity monitoring and speech enhancement from microphone arrays.

\bibliographystyle{IEEEbib}
\bibliography{main}
\end{document}